# Granularity Controlled Non-Saturating Linear Magneto-resistance in Topological Insulator Bi$_2$Te$_3$ Films

Z.H. Wang,[†] L. Yang,[†] X.J. Li,[†] X.T. Zhao,[†] H.L. Wang,[†] Z.D. Zhang [†,*] and Xuan P. A. Gao [‡,*]

[†]Shenyang National Laboratory for Materials Science, Institute of Metal Research, Chinese Academy of Sciences, 72 Wenhua Road, Shenyang 110016, People's Republic of China

[‡]Department of Physics, Case Western Reserve University, Cleveland, Ohio 44106, United States

[*] Email: (Z.D.Z) zdzhang@imr.ac.cn; (X.P.A.G.) xuan.gao@case.edu

## Abstract

**We report on the magneto-transport properties of chemical vapor deposition grown films of interconnected Bi$_2$Te$_3$ nanoplates. Similar to many other topological insulator (TI) materials, these granular Bi$_2$Te$_3$ films exhibit a linear magneto-resistance (LMR) effect which has received much recent attention. Studying samples with different degree of granularity, we find a universal correlation between the magnitude of the LMR and the average mobility ($<\mu>$) of the films over nearly two orders of magnitude change of $<\mu>$. The granularity controlled LMR effect here is attributed to the mobility fluctuation induced classical LMR according to the Parish-Littlewood theory (Nature 2003). These findings have implications to both the fundamental understanding and magneto-resistive device applications of TI and small bandgap semiconductor materials.**

Materials with large magneto-resistance (MR) have been of great interest in both fundamental research and device applications. It has been established for more than half a century that classical MR in most metals has a quadratic dependence on magnetic field (*B*) and tends to saturate in high fields. In some rare cases, a non-saturating linear MR may be found in metals with open Fermi surface.[1] In semiconductors with small or near-zero band gap, a variety of interesting MR phenomenon has been discovered over the last two decades. In doped silver chalcogenides $Ag_{2+\delta}Se$ and $Ag_{2+\delta}Te$, an anomalously large MR was observed, which depended linearly on magnetic field without any sign of saturation at fields as high as 60 T, over the temperatures range 4.5 to 300 K.[2,3] Moreover, tuning the bandgap and disorder in silver chalcogenides by pressure allowed enhancement of the LMR when the Hall resistivity changes sign and bands cross.[4] In addition to chalcogenides, polycrystalline narrow band-gap semiconductors such as InSb[5,6] and multi-layer graphene[7] were also found to exhibit LMR. Therefore, positive LMR appears to be a rather ubiquitous effect in small or near-zero band gap semiconductors, especially when sample has non-negligible granularity/inhomogeneity. Based on these two key features of experimental systems, many theoretical efforts have also put forward to explain the LMR.[8-15] Among the theoretical models, the so-called 'quantum MR' theory by Abrikosov is based on the readiness of zero-gap semiconductor being in the high field quantum limit where only one Landau level is occupied and a non-saturating MR is possible.[8,9] Emphasizing the importance of sample inhomogeneity and distorted current distribution, a phenomenological semiclassical random resistor network model by Parish and Littlewood was able to explain the linear MR using classical physics.[12,13] Similar to the idea of Parish-Littlewood, other models based on inhomogeneous conduction were also proposed.[14,15]

In the last few years, the rapid expansion of the emerging field of topological insulators (TIs) has again stimulated intensive research on the MR in these topological materials with non-trivial zero-gap Dirac-like surface states.[16-26] After the initial discovery of a LMR together with the two-dimensional (2D) quantum oscillations of MR from the surface states,[16,17] many further experiments confirmed the existence of LMR in TI materials.[18-26] These investigations on high quality $Bi_2Se_3$, $Bi_2Te_3$ films and individual nanoflakes have enabled observation of extremely large effect (e.g. over 600% increase in MR at room temperature in $Bi_2Te_3$ nanoflakes [19]), persistence of LMR to extreme conditions (e.g up to 60 Tesla in $Bi_2Te_3$ films[20]), and thickness or gate tunable LMR.[18, 22] Utilizing tilted magnetic field measurement, the linear MR of single crystalline TI in perpendicular field was attributed to the 2D gapless topological surface states and of quantum origin.[17-20] On the other hand, other MR analysis suggested that charge inhomogeneity and conductivity fluctuation are also important, despite the sample being high quality film or individual single crystalline flake.[21, 25, 26] For instance, when the transport is tuned from bulk to topological surface conduction by applying back gate, the strong enhancement of linear MR is accompanied by a strongly nonlinear Hall effect, alluding to the relevance of charge inhomogeneity.[25] Thus, up to date; it still remains unclear if the LMR in TI is due to the unique zero-gap nature of surface states or any physical/electronic inhomogeneity in sample. Besides elucidating the basic mechanism of LMR, it would also be interesting to find additional methods to control the LMR in TI.

Here, we report a study of $Bi_2Te_3$ films of interconnected nanoplates synthesized by CVD method. The non-saturating LMR was observed in these films with variable granularity or uniformity up to 14T magnetic field, the highest field available in our instrument. A close correlation between LMR, average mobility and the sample's uniformity was revealed over a broad range of parameter space (near two orders of magnitude change in mobility and LMR's magnitude). Our work on granular films of $Bi_2Te_3$ nanoplates provides a definite evidence for the relevance of sample's physical or structural inhomogeneity in the origin of LMR and offers a new route to control the magneto-resistive properties of TI materials.

$Bi_2Te_3$ films were grown by CVD method on semi-insulating Si substrates with size of ~1.5cm×1.5cm in 10% $H_2$/Ar carrier gas. The growth method is similar to our previous work on $Bi_2Te_3$ and $Bi_2Se_3$ nanomaterials[28, 29] except that here we focused on examining the high deposition temperature area of growth substrate where deposition is rich and $Bi_2Te_3$ nanoplates are interconnected into a film. Briefly, a high temperature tube furnace (Lindberg Blue M) and one inch diameter quartz tube were used as the reactor for the synthesis with accurate control of temperature and gas flow rate. 99.99% $Bi_2Te_3$ powder was used as precursor and placed at the center of the furnace at $500^oC$. The growth substrate was placed at a distance of 14-15cm away from the $Bi_2Te_3$ source. At first, maximum (240sccm) flow rate of $H_2$/Ar carrier gas was introduced after the system was pumped down to the base pressure of about 0.2Pa. When the temperature of the furnace reached to $500^oC$, the flow rate was decreased to 40sccm. The pressure was kept at 30-40Pa for 5min to deposit the $Bi_2Te_3$ film. After that the furnace was cooled naturally down to room temperature and dark grey deposition composed of $Bi_2Te_3$ were found on the substrates. Generally, the $Bi_2Te_3$ deposition on the ~1.5cm long growth wafer consists of film of interconnected $Bi_2Te_3$ nanoplates in the high temperature region (close to the center of furnace). The nanoplate density of film gradually decreases as the deposition temperature deceases (for positions on the growth wafer that are farther from the center of furnace). More than ten samples were prepared with different thickness and granular size of nanoplates, but in the following we focus on discussing representative data collected on four samples noted as sample 1, 2, 3 and 4 respectively. The structural and morphological characterizations were performed by scanning electron microscope (SEM) and energy dispersive X-ray was used to confirm the stoichiometry of $Bi_2Te_3$. To ensure the consistent growth condition over the measured film, rectangle shaped samples with length about 1cm and width of only 3~4mm were cut from growth wafer for transport measurement. The transport properties were investigated by Quantum Design Physical Property Measurement System with a 14T magnet and temperature range from 300 to 2 K. The longitudinal and Hall resistance versus temperature and magnetic field curves were collected with a standard six terminals Hall bar configuration.

The temperature dependent resistance at *B*=0 for sample 1, 2, 3 and 4 are shown in Fig. 1(a), representing low resistivity, medium resistivity, nearly semiconducting and semiconducting samples. For each type of behavior, we have measured at least two samples and observed consistent behavior in LMR described later. For the first three categories of samples, apart from the initial rise of $R$ at ~300K due to carrier freeze out, $R(T)$ is metallic from $T$ = 2K to about 250 K, presumably because phonon scattering is weaker at low $T$ and that the unintentionally formed defects during the growth induce high carrier concentration and give rise to metallic bulk conduction, similar to other chalcogenide TI materials without intentional doping.[30-32] However, the $R(T)$ of sample 4 decreases with increasing temperature showing semiconductor behavior in most of the temperature range. We believe that this is not due to the shifting of Fermi level into bulk band gap, but rather the much stronger scatterings from disorder and porosity in this type of samples since carrier density is measured to be similar to other samples yet mobility is much lower in sample 4. Typical SEM images of these four $Bi_2Te_3$ films are shown in Fig. 1(b)-(e). As showing in SEM images, the films are formed by interconnected grains of $Bi_2Te_3$ nanoplates with different sizes. We also find vertically grown nanoplates (most time shown as bright white blade-like features inside the porous area in SEM image) in some samples but the interconnected nanoplates laying parallel to the sample surface should dominate the measured transport in our experiments. At a qualitative level, one can already see the interesting connection between sample's $R(T)$ and granularity/porosity: correlated with the increasing resistance, sample 1 to 4 consist of more and more loosely packed nanoplate networks. Moreover, sample 1, 2, and 3 have similar nanoplate grain size ~1 μm with increasingly less connectivity between plates while sample 4 is made of grains and holes with larger size and there are also more stacking between grains (additional SEM images shown in Fig. S1). As a way to quantify the granularity of film, we plot in Fig.1f the number of nanoplates counted over an area of $200\times200\mu m^2$ *vs*. the average mobility (to be discussed later) of the sample at $T$=2K. It can be seen that higher density of nanoplates in the film leads to lower resistivity and higher mobility, due to the improved connectivity of the network. The thickness of films can be estimated from the SEM images of sample taken from the side view (SEM images shown in Fig.S2). The layer thickness of films is typically about 100 nm.

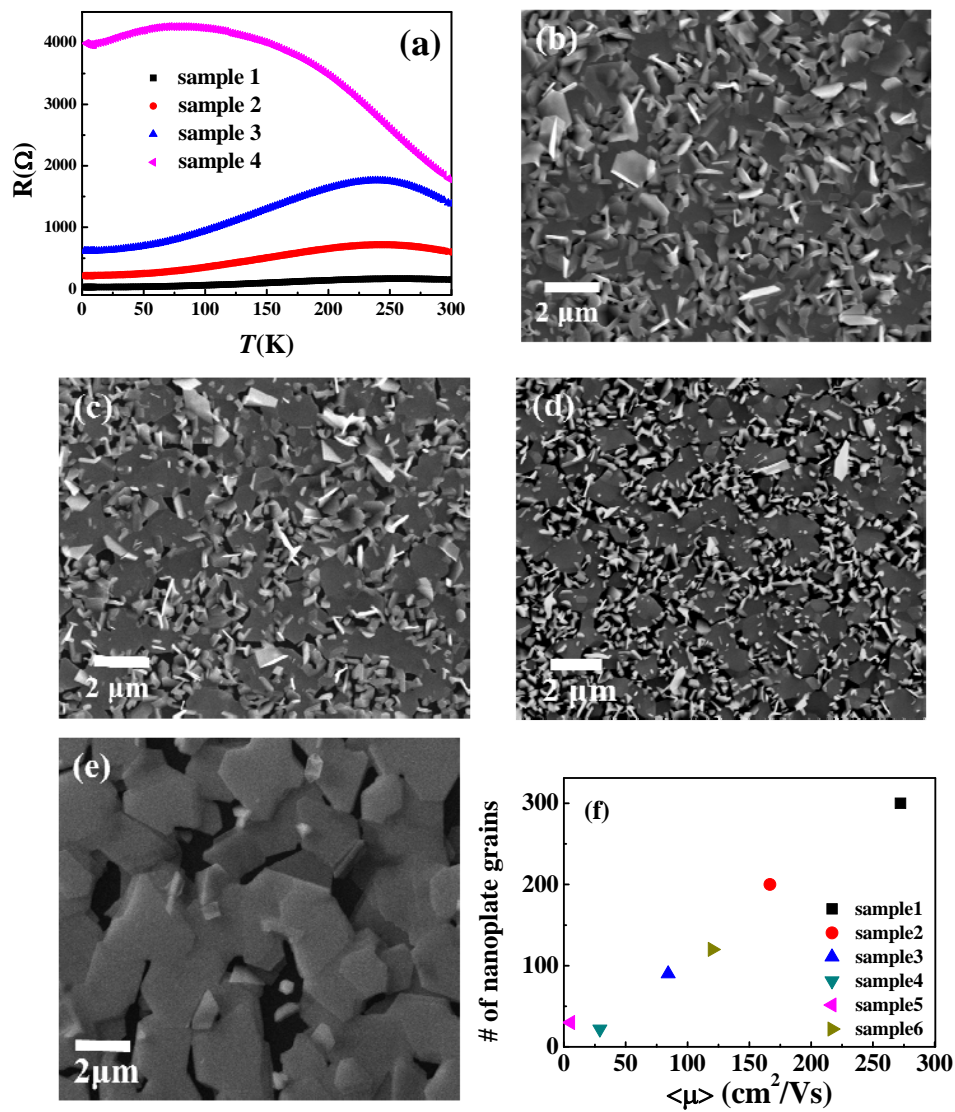

**Figure 1.** (a) The resistance plotted as a function of temperature for four representative samples. (b)-(e) SEM images of sample 1, 2, 3 and 4. (f) Number of nanoplate grains per 20 μm ×20μm area *vs.* the average mobility of the film at 2K.

Perpendicular magnetic field induced MR for sample 1, 2, 3, and 4 were measured over $T$= 2 - 300K and $B$= -14T to +14T. Since the normalized MR, $\Delta R(B)/R(0)$ highlights the intrinsic properties, we present $\Delta R(B)/R(0)$ in Fig.2. The raw $R(B)$ data are shown in Fig. S3. Similar to previous work [17] on single crystal TI materials, all the granular $Bi_2Te_3$ nanoplate film samples show LMR effect which decreases with increasing temperature. Also, $\Delta R(B)/R(0)$ first shows a quadratic growth below certain threshold field, and then transforms into a linearly rising behavior with increasing magnetic field without sign of saturation. Strikingly, we observe that there is a marked difference in the magnitude of LMR between samples. In sample 1 with most densely packed nanoplates, a large LMR (~450% at 14T) was obtained at low temperatures. However, as the film became less densely packed, $\Delta R(B)/R(0)$ decreases until the maximal $\Delta R(B)/R(0)$ in sample 4 is merely one tenth of that in sample 1. In most of previous studies of LMR in TIs, single crystal samples were measured and generally the large MR is thought to be associated with the high quality of sample. Thus it is striking that such clear and strong LMR was seen in our granular $Bi_2Te_3$ films of nanoplates. We take this as a clear and direct indication that inhomogeneity or fluctuations in TI samples can indeed generate large LMR, as various theories predicted. [12,13] Interestingly, in sample 4 which contains the largest nanoplate grains and weakest classical LMR effect, a sharp resistance dip in the $\Delta R(B)/R(0)$ curve around $B$=0 appears at low temperature, manifesting the weak antilocalization (WAL) effect. [28, 33]

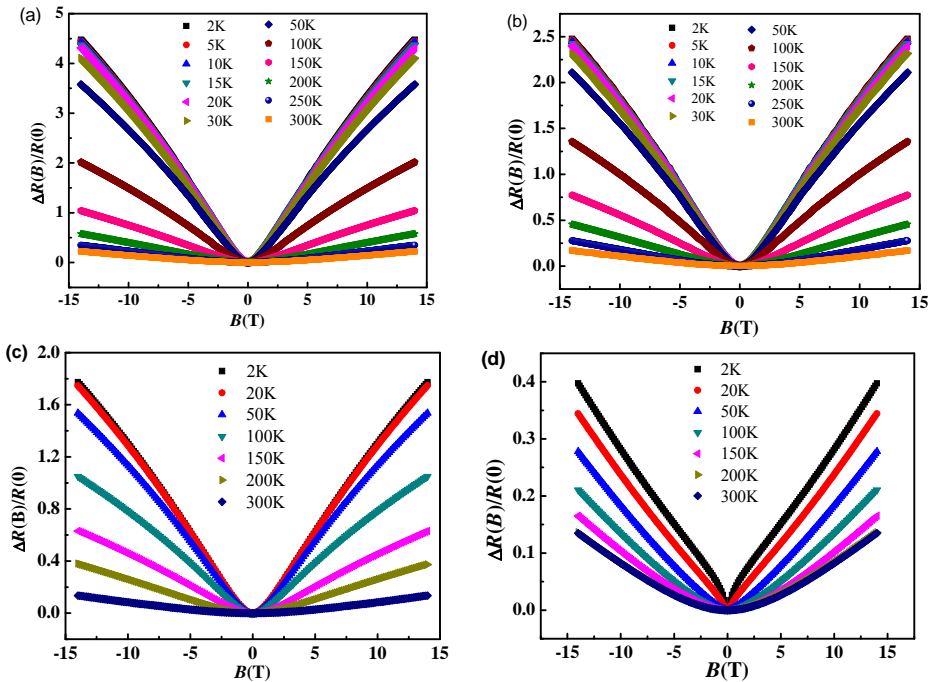

**Figure 2.** Magneto-resistance $\Delta R(B)/R(0)$ (defined as $[R(B)-R(0)]/R(0)$) as a function of magnetic field at different temperatures for sample1, 2, 3 and 4 (a-d).

Figure 3 shows the Hall resistance $R_{xy}$ as functions of $B$ at various temperatures. The sign of $R_{xy}$ indicates n-type conduction and a strong temperature dependence is seen in all samples. As shown in Fig. 3(a)–(c) and Fig. S4. The corresponding shape of $R_{xy}$ vs. $B$ changes from linear to non-linear with decreasing temperature, suggesting a change from one-band band transport to two (or multiple) band transport due to coexisting surface and bulk channels of carriers,[34,35] and/or electron and hole puddles.[25] Due to the complex structure of samples, we avoid fitting of $R_{xy}$ data to multi-band transport model with many parameters. Instead, we use the limiting high field slope of $R_{xy}(B)$, i.e. $dR_{xy}/dB$ at the highest $B$ to extract $n_{total}$, the total density of electrons from all bands since in the multiband transport model the asymptotic behavior of $dR_{xy}/dB$ is only determined by the sum of carrier densities from all the bands, independent of each band's mobility and make the fitting of $n_{total}$ a reliable one parameter fit. We obtain $n_{total}$ on the order of $10^{19}/cm^3$ for all the samples (Fig. 3(d)). Combining $n_{total}$ and the zero field resistance value, we extract the average mobility $<\mu>$ which is an important parameter in the theories of LMR based on mobility or conductivity fluctuations. We will next discuss the relations between $<\mu>$, LMR and granularity.

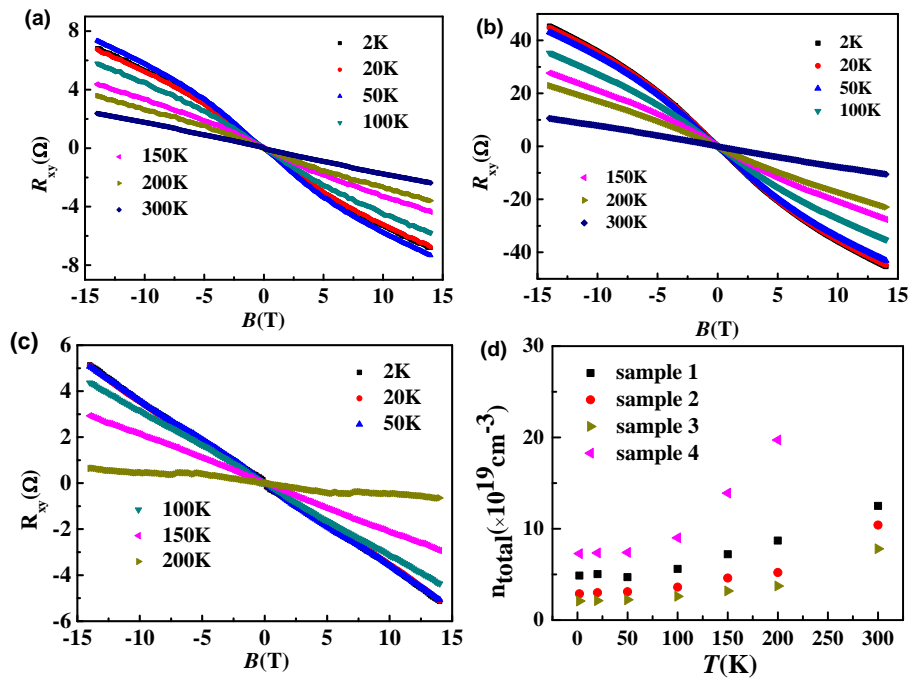

**Figure 3.** (a)-(c) Hall resistance data of sample 1, 3 and 4. (d) The total carrier concentration of different samples as a function of temperature.

The classical LMR found in inhomogeneous systems with strong disorder has been explained using phenomenological model by Parish and Littlewood.[12,13] The core of this model is a two dimensional network that consists of interconnected resistors in a random manner as such distorted current paths arise from disorder-induced inhomogeneity and macroscopic variations. In the numerical simulation, the Hall voltage associated with each resistor element could be parallel to the overall external voltage applied to the sample and thus induce a LMR.[12,13] Given that our samples are made of inter-connected $Bi_2Te_3$ nanoplate networks, they are an excellent system to test the Parish-Littlewood model. As shown in Fig. 4, the mobility and $\Delta R(B)/R(0)$ follow each other closely as temperature varies. Putting all the $\Delta R(B)/R(0)$ data for all the samples with different granularity at different temperatures together in the same graph, we found an remarkable correlation between $<\mu>$ and $\Delta R(B)/R(0)$ over wide range of $<\mu>$ (~5-300 $cm^2$/Vs), as shown in Fig.4d using $B$=14T as an example. In the Parish-Littlewood model, $\Delta R(B)/R(0) \propto <\mu>$ when $\Delta\mu/<\mu>$ is less than 1. So it appears that our data are consistent with the mobility fluctuation model in the regime that mobility disorder width $\Delta\mu$ is less than the average mobility itself. Here we caution that since $\mu B$ is the key scale in the scattering of orbiting electrons in magnetic fields, a generic correlation between mobility and MR is always expected for materials [12,14] and thus cannot be taken alone as the justification for the existence of mobility fluctuations. However, with various degree of granularity in our $Bi_2Te_3$ nanoplate films, we have confidence in that changes in $<\mu>$ are originated from varied granularity therefore so as the strength of LMR. Interestingly, in our system, increased density of nanoplates in film improves both the connectivity of the nanoplate network as well as the mobility and LMR's strength. Note that $\Delta R(B)/R(0)$ is predicted to be proportional to $\Delta\mu$ in the strong fluctuation regime of ($\Delta\mu/<\mu> > 1$),[12] such regime might be realized in sample 4 which has the smallest $<\mu>$ and $\Delta R(B)/R(0)$ larger than the generally linear correlation line between $<\mu>$ and $\Delta R(B)/R(0)$ in other samples with higher values of $<\mu>$.

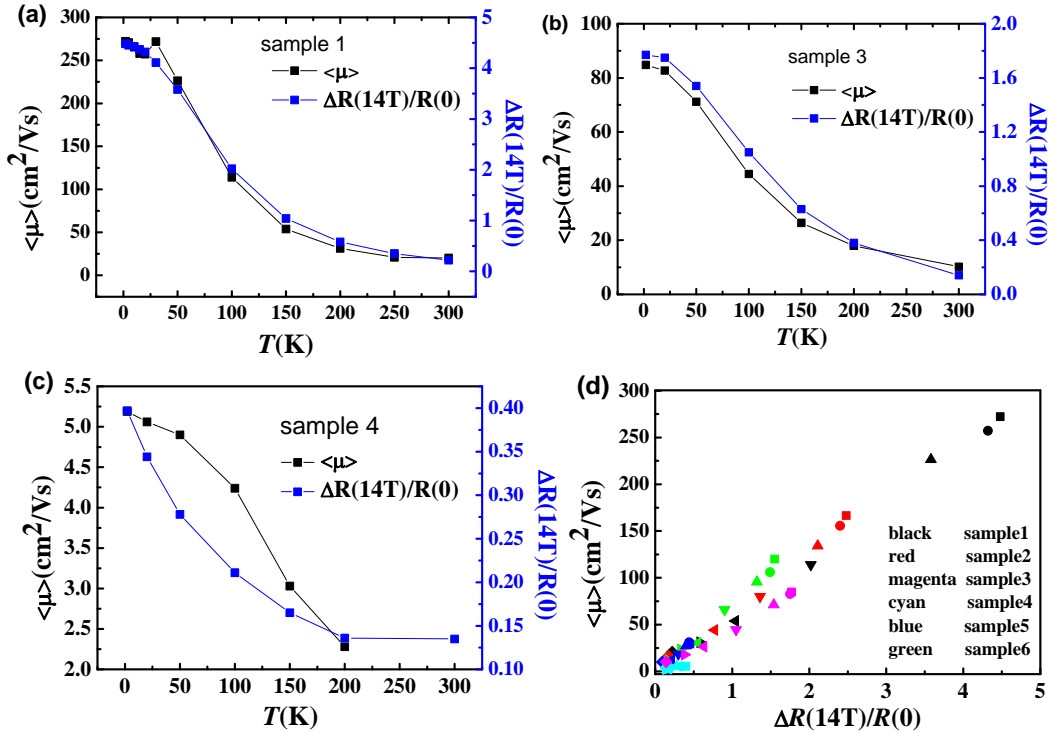

**Figure 4.** (a)-(c) Hall mobility and magneto-resistance at 14T plotted as a function of temperature for sample 1, 3, and 4. (d) The relation of average mobility and $\Delta R(B=14T)/R(B=0)$ for different samples with different granularity.

In summary, we have demonstrated that the giant linear and non-saturating MR in topological insulator $Bi_2Te_3$ can be tuned by adjusting the degree of granularity of inter-connected nanoplates in films prepared by CVD method. The proportional relation between mobility, sample's non-uniformity and LMR shows clear evidence for the phenomenological model of Parish-Littlewood, which attributes the LMR to large spatial fluctuations in the conductivity of the material. These findings shed light on both the basic mechanism of TI's MR as well as provide a new route towards controlling the MR of TIs.

## Acknowledgements

X. P. A. G acknowledges the NSF CAREER Award program (grant # DMR-1151534) for financial support of research at CWRU and the Lee Hsun Young Scientist award of IMR, Chinese Academy of Sciences. Z.D.Z acknowledges the National Natural Science Foundation of China with Grant No. 51331006.

**SYNOPSIS TOC**

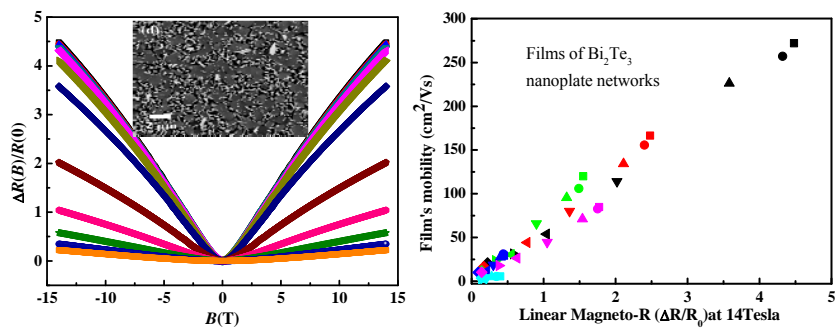